\documentclass[draft,twoside,leqno]{article}

\usepackage{amsmath,amsthm,amsfonts,amscd,mathrsfs,pifont} 
\usepackage{eucal}  
\usepackage{verbatim}     
\usepackage[dvips]{graphicx}
\usepackage{hyperref}

\newtheorem{thm}{Theorem}[section]
\newtheorem{cor}[thm]{Corollary}

\newtheorem{lemma}[thm]{Lemma}

\theoremstyle{definition}

\theoremstyle{remark}

\newcommand{\mding}[1]{\mbox{\ding{#1}}}

\newcommand{\BB}[1]{\ensuremath{\mathbb{#1}}}

\newcommand{\R}{\ensuremath{\BB{R}}}

\newcommand{\C}{\ensuremath{\BB{C}}}

\newcommand{\bs}{\ensuremath{\boldsymbol}}
\newcommand{\mf}{\ensuremath{\mathfrak}}

\newcommand{\la}{\ensuremath{\langle}}
\newcommand{\ra}{\ensuremath{\rangle}}

\DeclareMathOperator{\sgn}{sgn}

\DeclareMathOperator{\transpose}{T}
\DeclareMathOperator{\Tr}{Tr}

\DeclareMathOperator{\Pf}{Pf}

\numberwithin{equation}{section} 

\numberwithin{equation}{section}

\begin{document}
\title{\bf Averages over Ginibre's Ensemble of Random Real Matrices} 
\author{\sc Christopher D. Sinclair}
\maketitle

\begin{abstract}
We give a method for computing the ensemble average of 
multiplicative class functions over the Gaussian ensemble of real
asymmetric matrices.  These averages are expressed in terms
of the Pfaffian of Gram-like antisymmetric matrices formed with respect to a
skew-symmetric inner product related to the class function.
\end{abstract}

\section{Introduction}
In the 1960's Ginibre introduced three statistical ensembles of matrices
whose entries are chosen independently with Gaussian probability from (resp.)
$\R$, $\C$, and Hamilton's Quaternions \cite{MR0173726}.  These
ensembles are respectively labeled GinOE, GinUE and GinSE
in analogy with their Hermitian counterparts.  Ginibre
introduced a physical analogy between GinUE and two-dimensional
electrostatics, but gave no applications for the other two ensembles.
Since their introduction, many applications have been found for GinOE
and GinSE.

Here we report a method for determining
ensemble averages over GinOE of certain functions which are constant on
similarity (conjugacy) classes.  GinOE is the space of $N \times N$ real matrices $\R^{N
  \times N}$ together with the probability measure $\nu$ given   
by 
\[
\nu(S) := \mathcal{B}^{-1}_N \int_S \exp(-\Tr(X^{\transpose} X)/2) \, d\mu(X),
\]
where $\mu$ is Lebesgue measure on $\R^{N \times N}$, and
$\mathcal{B}_N = (2 \pi)^{N^2/2}$.  Our goal is to find
\[
\la \Psi \ra := \mathcal{B}_N^{-1} \int\limits_{\R^{N \times N}} \Psi(X)
\exp(-\Tr(X^{\transpose} X)/2) \, d\mu(X),
\]
where $\Psi: \R^{N \times N} \rightarrow \R$ is (i) constant on
similarity classes: $\Psi(A X A^{-1}) = \Psi(X)$ for all invertible $A
\in \R^{N \times N}$, and (ii) there exists a function $\phi: \C
\rightarrow \R$ such that if $D$ is a diagonal matrix with entries
$\gamma_1, \gamma_2, \ldots, \gamma_N$ then $\Psi(D) = \psi(\gamma_1)
\psi(\gamma_2) \cdots \psi(\gamma_N)$.  We remark that $\mu$-almost
every matrix in $\R^{N \times N}$ is similar to a diagonal matrix and 
consequently $\la \Psi \ra$ is uniquely determined by $\psi$. 

As an example, when $\psi(\gamma) = \gamma^n$, we are in the
well-studied situation $\Psi(X) = (\det X)^n$
\cite{nyquist-rice-riordan}\cite{cicuta-mehta}. 

Ginibre's original interest was the joint eigenvalue probability
density function (JPDF) and the $n$-point correlation functions of
GinUE, GinSE and GinOE. In the case of GinOE, Ginibre was only able to
report the JPDF in the restrictive case where all eigenvalues are
real: One difficulty in determining the full JPDF being that the space
of eigenvalues is naturally represented as a disjoint union indexed
over the possible numbers of real and complex conjugate pairs of
eigenvalues.  The full JPDF for GinOE was finally computed in the
1990's by Lehmann and Sommers \cite{lehmann-sommers}, and later
independently by Edelman \cite{edelman}.

Surprisingly, the formulation for the ensemble average of a
multiplicative class functions we present here is seemingly
independent of this decomposition of space.  Nonetheless, since
we will use the JPDF (which is dependent on this
decomposition) we introduce the details here.

Throughout the discussion $N$ will be fixed, and $(L, M)$ will
represent pairs of non-negative integers such that $L + 2M = N$. $L$
will represent the number of real eigenvalues 
and $M$ the number of complex conjugate pairs of eigenvalues.  The
symbols $\alpha$ and $\beta$ will represent real and (non-real)
complex eigenvalues respectively.  We also use $\lambda_L$ and $\lambda_{2M}$ to represent
Lebesgue measure on $\R^L$ and $\C^M$ respectively.  The space of all
possible eigenvalues of $N \times N$ matrices can then be identified
with the disjoint union
\[
\bigcup_{(L,M)} \R^L \times (\C \setminus \R)^M.
\]
It shall be convenient to write the partial JPDF's in terms of the
Vandermonde determinant.  Given $\boldsymbol{\gamma} \in \C^N$ we 
define $V^{\bs{\gamma}}$ to be the $N \times N$ Vandermonde matrix in
the coordinates of $\bs{\gamma}$.  (Superscripts on
matrices will indicate variables on which the entries are dependent).
The Vandermonde determinant is then given by
\begin{equation}
\label{eq:2}
\Delta(\bs{\gamma}) := \det V^{\bs{\gamma}} = \prod_{m < n} \gamma_n -
\gamma_m,  
\end{equation}
and given $\bs{\alpha} \in \R^L$ and $\bs{\beta} \in \C^M$ we define
\begin{equation}
\label{eq:13}
\Delta(\bs{\alpha}, \bs{\beta}) := \det V^{\bs{\gamma}}
\qquad where \qquad \bs{\gamma} := ( \overline{\beta_1}, \beta_1, \ldots,
\overline{\beta_M}, \beta_M, \alpha_1, \ldots, \alpha_L)
\end{equation}
The partial JPDF is then given by $P_{L,M}: \R^L \times (\C \setminus
\R)^M \rightarrow \R$ where
\begin{equation}
\label{eq:3}
  P_{L,M}(\boldsymbol{\alpha}, \boldsymbol{\beta}) = \mathcal{C}_N^{-1} \, \frac{|\Delta(\bs{\alpha}, \bs{\beta})|}{L! M!} 
 \prod_{\ell=1}^L e^{-\alpha_{\ell}^2/2}
\prod_{m=1}^M \mathrm{erfc}\left(\sqrt{2} \, |\mathrm{Im} \beta_m| \right) e^{-(\beta_m^2
  + \overline{\beta_m}^2)/2} 
,
\end{equation}
and 
\begin{equation}
\label{eq:10}
\mathcal{C}_N := 2^{N(N+1)/4} \prod_{n=1}^N \Gamma(n/2).
\end{equation}
This defines the full JPDF since the domains
of the partial JPDF's are disjoint.  The only essential
difference between the formulation for $P_{L,M}(\bs{\alpha},
\bs{\beta})$ presented here and that presented by Lehmann and Sommers
(and Edelman) is that they use the right hand side of (\ref{eq:2}). 

From (\ref{eq:3}) we can see the two main difficulties in the
computation of $\la
\Psi \ra$:  (i) the decomposition of the space of eigenvalues, and (ii)
the complicated nature of $|\Delta(\bs{\alpha}, \bs{\beta})|$.

\section{Statement of Results}

In order to state our results it shall be convenient to define
\[
\phi(\gamma) := \exp(-\gamma^2/2) \, \left\{\mathrm{erfc}(\sqrt{2} |\mathrm{Im}(
  \gamma)|)\right\}^{1/2} \, \psi(\gamma).
\]
We then define two skew-symmetric inner products,
\[
\la P, Q \ra_{\R} := \int_{\R^2}
\phi(\alpha_1) \phi(\alpha_2) \; P(\alpha_1) Q(\alpha_2) \; \sgn(\alpha_2
- \alpha_1) \, d\alpha_1 \, d\alpha_2, 
\]
and
\[
\la P, Q \ra_{\C} := -2 i \int_{\C} \phi(\beta) \phi(\overline{\beta}) \;
P(\overline{\beta}) Q(\beta) \; \sgn( \mathrm{Im}(\beta)) \, d\lambda_2(\beta) 
\]
where $\sgn(0) = 0$ and $\sgn(x) = x/|x|$.  These skew-symmetric inner
products also appear in another paper on the statistics of the
eigenvalues of matrices in GinOE by Kanzieper and Akemann
\cite{kanzieper-akemann}. 

\begin{thm}
\label{thm:1}
\sloppy
Let $J$ be the integer part of $(N+1)/2$.  If
$\mathbf{P} = \{P_1(\gamma), P_2(\gamma), \ldots, P_N(\gamma)\}
\subset \C[\gamma]$ is a set of monic polynomials with $\deg P_n
= n-1$, then, assuming $\la \Psi \ra$ exists,
\[
\la \Psi \ra = \mathcal{C}_N^{-1} \, \Pf U_{\mathbf{P}},
\]
the Pfaffian of $U_{\mathbf{P}}$, where $U_{\mathbf{P}}$ is the $2J
\times 2J$ antisymmetric matrix whose $j,k$  
entry is given by 
\begin{equation}
\label{eq:1}
U_{\mathbf{P}}[j,k] := \left\{
\begin{array}{ll}
\la P_j, P_k \ra_{\R} + \la P_j, P_k \ra_{\C}  & \quad \mbox{if } j, k \leq N, \\ 
{\displaystyle 
\sgn(k-j) \int_{\R} \phi(\alpha) \, P_{\min\{j,k\}}(\alpha) \, d\alpha
} & \quad \mbox{otherwise}, \\
\end{array}
\right.
\end{equation}
and $\mathcal{C}_N$ is given as in (\ref{eq:10}).
\end{thm}
Notice that when $N$ is even the first condition in
Equation~(\ref{eq:1}) always holds.  The asymmetry between even and
odd cases is due to the fact that the Pfaffian is not defined for odd
by odd matrices.

If $\mathbf{P}$ satisfies the
conditions of Theorem~\ref{thm:1}, we will say that $\mathbf{P}$ is a
{\it complete} set of monic polynomials.  When $N$ is even it is
sensible to use a complete family of monic polynomials which are {\it skew-orthogonal}.
\begin{cor}
\label{cor:1}
Suppose $N = 2J$, and let $\mathbf{Q} = \{Q_1, Q_2,
\ldots, Q_N\}$ be any complete family of monic polynomials specified by 
\[
\la Q_{2k-1}, Q_{2j} \ra = -\la Q_{2j}, Q_{2k - 1} \ra = \delta_{kj}
\mathfrak{M}_j \quad \mbox{and} \quad \la Q_{2j}, Q_{2k} \ra = \la
Q_{2j-1}, Q_{2k-1} \ra = 0,
\]
where $\la P, Q \ra = \la P, Q \ra_{\R} + \la P, Q \ra_{\C}$.  Then,
\[
\la \Psi \ra = \mathcal{C}_N^{-1} \, \prod_{j=1}^J \mathfrak{M}_j.
\]
\end{cor}
The quantities $\mf{M}_j$ are referred to as the {\it normalization(s)}
of $\mathbf{Q}$.  See \cite[Ch.~5]{mehta} or \cite[Ch.~5]{forrester}
for more about skew-orthogonal polynomials.

If $\psi$ satisfies an additional symmetry, then we may
write $\la \Psi \ra$ as a determinant
\begin{cor}
\label{cor:2}
Let $J$ be the integer part of $(N+1)/2$.    If $\psi(-\beta) =
\psi(\beta)$ for every $\beta \in \C$, and $\mathbf{P}$ is a 
complete family of monic polynomials in $\C[\gamma]$ such that $P_n$ is
even when $n-1$ is even and odd when $n-1$ is odd, then
\[
\la\Psi\ra = \mathcal{C}_N^{-1} \, \det A_{\mathbf{P}}
\]
where $A_{\mathbf{P}}$ is the $J \times J$ matrix whose $j,k$ entry is
given by
\[
A_{\mathbf{P}}[j,k] := U_{\mathbf{P}}[2j-1, 2k].
\]
\end{cor}

We reiterate the striking fact that these formulations of $\la \Psi
\ra$ are seemingly independent of the decomposition of the space of
eigenvalues into all the different possible numbers of real and
complex conjugate pairs of eigenvalues.  In fact, the decomposition of
the space of eigenvalues {\it does} enter into the statement of
Theorem~\ref{thm:1} --- it is the reason that the inner product in the
entries of $U_{\mathbf{P}}$ are {\it sums} of skew inner products.
One of these inner products introduces to $\la \Psi \ra$ contributions
from $\psi$ on real eigenvalues, while the other introduces
contributions from pairs of complex conjugate eigenvalues.

\section{Averages over GinUE}
The simplicity of Theorem~\ref{thm:1} and its corollaries can
(perhaps) be most appreciated when compared with the analogous results
for GinUE.   GinUE is the set $\C^{N \times N}$ together with the 
probability measure whose density is proportional to $\exp(
\Tr(Z^{\ast} Z)/2 )$.  The joint eigenvalue probability density
function was given by Ginibre as \begin{equation}
\label{eq:14}
P_N(\bs{\gamma}) := \mathcal{D}_N^{-1} \, | \Delta(\bs{\gamma}) |^2
\prod_{n=1}^N e^{-|\gamma_n|^2/2},
\qquad \mbox{where} \qquad
\mathcal{D}_N := (2\pi)^{N} \prod_{n=1}^N \Gamma(n+1).
\end{equation}

We will denote the ensemble average of $\Psi$
over GinUE by $\{ \Psi \}$, and we define the inner product,
\[
\la P | Q \ra := \int_{\C} e^{-|\gamma|^2/2} \, \psi(\gamma) \,
\overline{P(\gamma)} Q(\gamma) \, d\lambda_2(\gamma). 
\]
\begin{thm}
\label{thm:2}
Let $\mathbf{P}$ be any complete set of monic polynomials.  Then,
assuming $\{ \Psi \}$ exists,
\[
\{\Psi\} = \mathcal{D}_N^{-1} \det W_{\mathbf{P}},
\]
where $W_{\mathbf{P}}$ is the $N \times N$ symmetric matrix whose $j,k$
  entry is given by $W_{\mathbf{P}}[j,k] = \la P_j | P_k \ra,$
and $\mathcal{D}_N$ is given as in (\ref{eq:14}).
\end{thm}

\begin{cor}
Let $\mathbf{Q}$ be the complete family of monic polynomials which
are orthogonal with respect to $\la \cdot | \cdot \ra$, then
\[
\{ \Psi \} = \mathcal{D}_N^{-1} \prod_{n=1}^N \la Q_n | Q_n \ra.
\]
\end{cor}

The proof of Theorem~\ref{thm:2} is a straightforward exercise using
well-known techniques of random matrix theory (apply the JPDF to the
definition of $\{ \Psi \}$, expand $\Delta(\bs{\gamma})$ as a sum over
the symmetric group, apply Fubini's Theorem and simplify).  A similar
computation (in the domain of heights of polynomials) is done in
\cite{sinclair}.  The proof of \ref{thm:1} on the other hand is much
more difficult due to the previous mentioned complications.  The
analogy between \ref{thm:1} and \ref{thm:2} is made all the more
striking when comparing the differences in difficulties of their
proofs.  Clearly there is a deeper phenomenon at work.

\section{The Proof of Theorem~\ref{thm:1}}
The proof of Theorem~\ref{thm:1} is based on, indeed almost identical
to, another computation by the author in the study of heights of
polynomials with integer coefficients \cite{sinclair-2005}.  The
connection between that computation and the one presented here is that
the Jacobian of the change of variables from the roots to the
coefficients of a monic degree $N$ real polynomial with $L$ real roots
and $M$ pairs of complex conjugate roots is a constant times
$|\Delta(\bs{\alpha}, \bs{\beta})|/L!M!$.  Since the audiences of these two
results are likely disjoint many of the details are presented here,
though the proof of many formulas which are purely combinatorial will
only be referenced. 

From (\ref{eq:3}) we see,
\begin{equation}
\label{eq:5}
\la \Psi \ra = \mathcal{C}_N^{-1} \sum_{(L,M)} \frac{1}{L! M!}
\int\limits_{\R^L \times \C^M}  \left\{\prod_{\ell=1}^L \phi(\alpha)
\prod_{m=1}^M  \phi(\beta) \phi(\overline{\beta}) \right\}
 | \Delta(\bs{\alpha}, \bs{\beta}) | \,
d\lambda_L(\bs{\alpha}) \, d\lambda_{2M}(\bs{\beta}),
\end{equation}
Next, for each pair $(L,M)$ we use the Laplace expansion in order to
expand the Vandermonde determinant.  Since the first $2M$ columns of
$V^{\bs{\alpha},   \bs{\beta}}$ depend only on $\bs{\beta}$, while the
remaining columns depend only on $\bs{\alpha}$, we will expand 
$\Delta(\bs{\alpha}, \bs{\beta})$ via $2M \times 2M$ and $L \times L$
minors.  

\subsection{Notation for Minors}
For each $K \leq N$ we define $\mf{I}_K^N$ to be the set of increasing
functions from $\{1,2,\ldots, K\}$ to $\{1,2\ldots, N\}$.  That is,
\[
\mf{I}_K^N := \big\{ \{1,2,\ldots,K\}
\stackrel{\mathfrak{t}}{\longrightarrow} \{1,2,\ldots, N\} \;\; : \;\;
\mathfrak{t}(1) < \mathfrak{t}(2) < \cdots < \mathfrak{t}(K) \big\}.
\]
Associated to each $\mathfrak{t} \in \mf{I}_K^N$ there exists a unique
$\mathfrak{t}' \in \mf{I}_{N-K}^N$ such that the images of $\mathfrak{t}$
and $\mathfrak{t}'$ are disjoint.  Each $\mathfrak{t} \in \mf{I}_K^N$
induces a unique permutation in $\iota_{\mf{t}} \in S_N$ by specifying
that 
\[
\iota_{\mf{t}}(n) := \left\{
\begin{array}{ll}
\mf{t}(n) & \mbox{if} \quad 1 \leq n \leq K, \\ 
\mf{t}'(n-K) & \mbox{if} \quad K < n \leq N.
\end{array}
\right.
\]
We define the {\it sign} of $\mf{t}$ by setting $\sgn(\mf{t}) :=
\sgn(\iota_{\mf{t}})$.  The identity map in $\mf{I}_K^N$ is denoted by
$\mf{i}$.  

Given an $N \times N$ matrix $W$ and $\mf{u}, \mf{t} \in \mf{I}_K^N$,
define $W_{\mf{u},\mf{t}}$ to be the $K \times K$ minor whose $j,k$
entry is given by $W_{\mf{u}, \mf{t}}[j,k] = W[\mf{u}(j),
\mf{t}(k)]$.  The complimentary minor is given by $W_{\mf{u}',
  \mf{t}'}$.  As an example of the utility of this
notation, the Laplace expansion of the determinant can be written as
\begin{equation}
\label{eq:4}
\det W = \sgn(\mf{u}) \sum_{\mf{t} \in \mf{I}_K^N} \sgn(\mf{t}) \det
W_{\mf{u}, \mf{t}} \cdot \det W_{\mf{u}', \mf{t}'},
\end{equation}
where $\mf{u}$ is any fixed element of $\mf{I}_K^N$.  We will also use
the abbreviated notation $W_{\mf{u}}$ for $W_{\mf{u}, \mf{u}}$; this is
useful notation for working with Pfaffians since if $W$ is an
antisymmetric matrix then minors of the form $W_{\mf{u}}$ are also
antisymmetric.  

We recall the definition of the Pfaffian here.  If $N = 2J $ and $U$ is an $N \times N$ antisymmetric
matrix, then the Pfaffian of $U$ is given by
\begin{equation}
\label{eq:52}
\Pf U = \frac{1}{2^J J!} \sum_{\tau \in S_N} \sgn(\tau) \prod_{j=1}^J
U[\tau(2j-1), \tau(2j)],  
\end{equation}
where $S_N$ is the $N$th symmetric group.  We will also use the formula
\begin{equation}
\label{eq:8}
\Pf U = \frac{1}{J!} \sum_{\sigma \in \Pi_{2J}} \sgn(\sigma)
\prod_{j=1}^J U[\tau(2j-1), \tau(2j)],  
\end{equation}
where $\Pi_{2J}$ is the subset of $S_{2J}$ composed of those
$\sigma$ with $\sigma(2j) > \sigma(2j - 1)$ for $j=1,\ldots,J$.  The Pfaffian is related to
the determinant by the formula $\det U = (\Pf U)^2$ (see for instance 
\cite[Appendix: Pfaffians]{MR900587}). 

If $U = R + C$ where $R$ and $C$ are antisymmetric $2J \times 2J$
matrices, then $\Pf U$ has an expression in terms of the minors of $R$
and $C$ \cite[Lemma~8.7]{sinclair-2005}.
\begin{equation}
\label{eq:9}
\Pf U = \sum_{M=0}^J \sum_{\mf{u} \in \mf{I}_{2M}^{2J}} \sgn(\mf{u})  \Pf
R_{\mf{u}'} \cdot \Pf C_{\mf{u}}.
\end{equation}
This will be useful since ultimately we intend to express $\la \Psi
\ra$ as the Pfaffian of a sum of matrices.

\subsection{Steps in the Proof}
\begin{lemma}
\label{lemma:1}
Let $\bs{\gamma} \in \C^N$ be given as in (\ref{eq:13}) and let
$W^{\bs{\alpha}, \bs{\beta}}$ be the $N \times N$ matrix whose $j,k$
entry is given by
\[
W^{\bs{\alpha}, \bs{\beta}}[j,k] := P_k(\gamma_j).
\]
Then, if $\mf{i} \in \mf{I}_{2M}^N$ is the identity map on $\{1,2,\ldots, 2M\}$,
\[
\left| \Delta({\bs{\alpha}, \bs{\beta}}) \right| = \sum_{\mf{t}\in
  \mf{I}_{2M}^N} \sgn(\mf{t}) \left\{ \det W^{\bs{\beta}}_{\mf{i}, \mf{t}}
  (-i)^M \prod_{m=1}^M \sgn \Im(\beta_m) \right\} \Bigg\{ \det
W_{\mf{i}', \mf{t}'}^{\bs{\alpha}} \prod_{j < k} \sgn(\alpha_k -
\alpha_j) \Bigg\} ,
\]
where as suggested by the notation, the minors $W_{\mf{i},
  \mf{t}}^{\bs{\beta}}$ and $W^{\bs{\alpha}}_{\mf{i}',
  \mf{t}'}$ of $W^{\bs{\alpha}, \bs{\beta}}$ are dependent only on
$\bs{\beta}$ and $\bs{\alpha}$ respectively.
\end{lemma}
Using Lemma~\ref{lemma:1}, Equation~(\ref{eq:5}) becomes
\begin{eqnarray*}
\lefteqn{\la \Psi \ra = \mathcal{C}_N^{-1} \sum_{(L,M)} \frac{1}{M! L!} \sum_{\mf{t} \in
    \mf{I}_{2M}^N} 
\sgn(\mf{t}) \int\limits_{\R^L} \int\limits_{\C^M} \left\{\prod_{\ell=1}^L
\phi(\alpha_{\ell})\right\}\left\{ \prod_{m=1}^M \phi(\beta_m)
\phi(\overline{\beta_m})\right\}} \\ && \times
 \Bigg\{
\det W_{\mf{i}', \mf{t}'}^{\bs{\alpha}} \prod_{j < k}
\sgn(\alpha_k - \alpha_j) 
\Bigg\}
 \left\{ 
\det W^{\bs{\beta}}_{\mf{i}, \mf{t}} (-i)^M \prod_{m=1}^M \sgn \Im(\beta_m)
\right\} \, d\lambda_L(\bs{\alpha}) \, d\lambda_{2M}(\bs{\beta}),
\end{eqnarray*}and Fubini's Theorem yields
\begin{eqnarray}
\lefteqn{
\la \Psi \ra = \mathcal{C}_N^{-1} \sum_{(L,M)} \sum_{\mf{t} \in \mf{I}_{2M}^N}
\sgn(\mf{t}) 
\frac{1}{L!} 
\int\limits_{\R^L} \left\{\prod_{\ell=1}^L \phi(\alpha_{\ell})
\prod_{j<k} \sgn(\alpha_k - \alpha_j)\right\} \, \det
W^{\bs{\alpha}}_{\mf{i}', \mf{t}'} \; d\lambda_{L}(\bs{\alpha})} \nonumber \\
&& \hspace{2.7cm} \times \frac{(-i)^M}{M!} \int\limits_{\C^M}
\left\{\prod_{m=1}^M \phi(\beta_m) \phi(\overline{\beta_m}) \sgn
\Im(\beta_m)\right\} \, \det W^{\bs{\beta}}_{\mf{i},\mf{t}} \;
d\lambda_{2M}(\bs{\beta}). \label{eq:7}
\end{eqnarray}
We remark that by using Fubini's Theorem we are assuming that the
average $\la \Psi \ra$ actually exists. 

Next, let $K$ be the integer part of $(L+1)/2$.  Define $T^{\bs{\alpha}}$ to
be the $2K \times 2K$ antisymmetric matrix whose $j,k$ entry is given by
\begin{equation}
\label{eq:11}
T^{\bs{\alpha}}[j,k] := \left\{
\begin{array}{ll}
\sgn(\alpha_k - \alpha_j) & \quad \mbox{if} \quad j,k < L+1, \\
\sgn(k-j)  & \quad \mbox{otherwise.}
\end{array}
\right.
\end{equation}
Then, it is well known that,
\begin{equation}
\label{eq:6}
\prod_{1 \leq j < k \leq L} \sgn(\alpha_k - \alpha_j) = \Pf T^{\bs{\alpha}},
\end{equation}
(see \cite[Ch.5]{forrester}, or \cite[Lemma 8.4]{sinclair-2005}).
It is worth remarking that when $L$ is even then the first condition
defining $T^{\bs{\alpha}}$ is always in force.  Since the Pfaffian is
only defined for even rank antisymmetric matrices, the second
condition is used when $L$ is odd to bootstrap a $2K \times 2K$ antisymmetric
matrix from an $L \times L$ matrix. 

Substituting (\ref{eq:6}) into (\ref{eq:7}) we see
\begin{eqnarray}
\lefteqn{
\la \Psi \ra = \mathcal{C}_N^{-1} \sum_{(L,M)} \sum_{\mf{t} \in \mf{I}_{2M}^N}
\sgn(\mf{t}) \frac{1}{L!} 
\int\limits_{\R^L} \left\{\prod_{\ell=1}^L \phi(\alpha_{\ell}) \right\}
\, \Pf T^{\bs{\alpha}} \cdot \det
W^{\bs{\alpha}}_{\mf{i}', \mf{t}'} \; d\lambda_{L}(\bs{\alpha})} \nonumber \\
&& \hspace{2cm} \times \frac{(-i)^M}{M!} \int\limits_{\C^M}
\left\{\prod_{m=1}^M \phi(\beta_m) \phi(\overline{\beta_m}) \sgn
\Im(\beta_m)\right\} \, \det W^{\bs{\beta}}_{\mf{i},\mf{t}} \;
d\lambda_{2M}(\bs{\beta}). \label{eq:34}
\end{eqnarray}Now, let $J$ be the integer part of $(N+1)/2$.  It is necessary for our calculations to replace the $\mf{t} \in
\mf{I}_{2M}^N$ with elements of $\mf{I}_{2M}^{2J}$.  Each $\mf{t} \in \mf{I}_{2M}^N$
induces a unique $\mf{t}_{\circ} \in \mf{I}_{2M}^{2J}$ by setting
$\mf{t} = \mf{t}_{\circ}$.  Notice that $\mf{t}'$ and
$\mf{t}_{\circ}'$ differ in the fact that if $N \neq 2J$ then
$\mf{t}'_{\circ}(2J-2M) = 2J$.  Clearly, $\sgn(\mf{t}_{\circ}) =
\sgn(\mf{t})$.
\begin{lemma}
\label{lemma:7}
Let $R$ be the $2J \times 2J$ matrix whose $j,k$ entry is given by
\[
R[j,k] := \left\{
\begin{array}{ll}
\la P_j, P_k \ra_{\R} & \quad \mbox{if} \quad j,k < N+1 \\
{\displaystyle \sgn(k-j) \int_{\R} \phi(\alpha) \, P_{\min\{j,k\}}(\alpha) \, d\alpha}  & \quad \mbox{otherwise},
\end{array}
\right.
\]
and suppose $\mf{t} \in \mf{I}_{2M}^N$.  Then,
\[
\frac{1}{L!}
\int\limits_{\R^L} \left\{\prod_{\ell=1}^L \phi(\alpha_{\ell}) \right\}
\, \Pf T^{\bs{\alpha}} \, \det
W^{\bs{\alpha}}_{\mf{i}', \mf{t}'} \; d\lambda_{L}(\bs{\alpha}) = \Pf R_{\mf{t_{\circ}'}}.
\]
\end{lemma}
When $N$ is odd and $\mf{t} \in \mf{I}_{2M}^{N}$ then
$R_{\mf{t}'}$ is an odd by odd matrix.  The introduction of
$\mf{t}_{\circ}$ is useful since the Pfaffian of $R_{\mf{t}_{\circ}'}$
is defined.
\begin{lemma}
\label{lemma:8}
Let $C$ be the $2J \times 2J$ matrix whose $j,k$ entry is given by
\[
C[j,k] := \left\{
\begin{array}{ll}
\la P_j, P_k \ra_{\C} & \quad \mbox{if} \quad j,k < N+1 \\
0 & \quad \mbox{otherwise},
\end{array}
\right.
\]
and suppose $\mf{t} \in \mf{I}_{2M}^{N}$.  Then,
\[
\frac{(-i)^M}{M!} \int\limits_{\C^M}
\left\{\prod_{m=1}^M \phi(\beta_m) \phi(\overline{\beta_m}) \sgn
\Im(\beta_m)\right\} \, \det W^{\bs{\beta}}_{\mf{i},\mf{t}} \;
d\lambda_{2M}(\bs{\beta}) = \Pf C_{\mf{t_{\circ}}}.
\]
\end{lemma}
Using Lemma~\ref{lemma:7} and Lemma~\ref{lemma:8} we may rewrite
(\ref{eq:34}) as
\begin{equation}
\label{eq:35}
\la \Psi \ra = \mathcal{C}_N^{-1}
\sum_{(L,M)} \sum_{\mf{t} \in \mf{I}_{2M}^N} \sgn(\mf{t}_{\circ}) \Pf R_{\mf{t}_{\circ}'} \cdot \Pf C_{\mf{t}_{\circ}}.
\end{equation}
If $\mf{u} \in \mf{I}_{2M}^{2J}$ then either $2J$ is in the image of
$\mf{u}$ or $2J$ is in the image of $\mf{u}'$.  Notice that if $2J$ is
in the image of $\mf{u}$ then $\Pf C_{\mf{u}} = 0$.  If $2J$ is in the
image of $\mf{u}'$ then $\mf{u}'(2J-2M) = 2J$ and hence $\mf{u} =
\mf{t}_{\circ}$ for some $\mf{t} \in \mf{I}^N_{2M}$.  Thus we may
replace the sum over $\mf{I}_{2M}^N$ in (\ref{eq:35}) with a sum over
$\mf{I}_{2M}^{2J}$.  Consequently,
\begin{align}
\la \Psi \ra = \; & 
\mathcal{C}_N^{-1} \sum_{(L,M)} \sum_{\mf{u} \in \mf{I}_{2M}^{2J}} \sgn(\mf{u}) \Pf
R_{\mf{u}'} \cdot \Pf C_{\mf{u}} \nonumber \\
= \; & \mathcal{C}_N^{-1} \sum_{M=0}^J \sum_{\mf{u} \in \mf{I}_{2M}^{2J}} \sgn(\mf{u}) \Pf
R_{\mf{u}'} \cdot \Pf C_{\mf{u}},
\end{align}
where the second equation follows since the summand has been made to
be independent of $L$.  From (\ref{eq:9}) we see that $\la \Psi \ra =
\mathcal{C}_N^{-1} \Pf(R +
C)$.  Consequently, by (\ref{eq:1}), $\la \Psi \ra = \mathcal{C}_N^{-1}
\, \Pf U_{\mathbf{P}}$.

\subsection{The Proof of Lemma~\ref{lemma:1}}
Applying (\ref{eq:2}) to (\ref{eq:13}),
\begin{eqnarray*}
\lefteqn{\Delta({\boldsymbol{\alpha}, \boldsymbol{\beta})} =
  \left\{\prod_{j < 
  k}(\alpha_k - \alpha_j) \right\} \prod_{l=1}^L \prod_{m=1}^M \left|\beta_m
- \alpha_l \right|^2} \quad \quad \\
&\times& \left\{
\prod_{m < n} \left|\beta_n - \beta_m \right|^2 \left|\beta_n -
  \overline{\beta_m} \right|^2 
\right\} \prod_{m=1}^M  2 i \Im(\beta_m). 
\end{eqnarray*}
And hence,
\begin{equation}
\left|\Delta(\bs{\alpha}, \bs{\beta})\right| = (-i)^M \left\{
\prod_{j < k} \sgn(\alpha_k - \alpha_j) \prod_{m=1}^M \sgn \Im(\beta_m)
\right\} \Delta({\bs{\alpha}, \bs{\beta}}).
\label{eq:40}
\end{equation}
We may replace the monomials in the Vandermonde matrix with any
complete family of monic polynomials without changing its
determinant.  That is, $\Delta({\bs{\alpha}, \bs{\beta}}) = \det
W^{\bs{\alpha}, \bs{\beta}}$.  Using the Laplace expansion of the
determinant (\ref{eq:4}) with $\mf{u} = \mf{i} \in
\mf{I}_{2M}^N$, we see that
\[
\det W^{\bs{\alpha}, \bs{\beta}} = \sum_{\mf{t} \in \mf{I}_{2M}^N}
\sgn(\mf{t}) \det W^{\bs{\alpha}, \bs{\beta}}_{\mf{i},\mf{t}} \cdot \det
W^{\bs{\alpha}, \bs{\beta}}_{\mf{i}', \mf{t}'}.
\]
Notice that the minors of the form $W^{\bs{\alpha},
  \bs{\beta}}_{\mf{i}, \mf{t}}$ consists of elements from the first
$2M$ columns of $W^{\bs{\alpha}, \bs{\beta}}$.  These columns are not
dependent on $\bs{\alpha}$ and thus we may write these minors as
$W^{\bs{\beta}}_{\mf{i}, \mf{t}}$.  Similarly we may write the minors
of the form $W^{\bs{\alpha}, \bs{\beta}}_{\mf{i}', \mf{t}'}$ as
$W^{\bs{\alpha}}_{\mf{i}', \mf{t}'}$.  It follows that
\begin{equation}
\label{eq:42}
\det V^{\bs{\alpha}, \bs{\beta}} = \sum_{\mf{t} \in \mf{I}_{2M}^N}
\sgn(\mf{t}) \det W^{\bs{\beta}}_{\mf{i},\mf{t}} \cdot \det
W^{\bs{\alpha}}_{\mf{i}', \mf{t}'},
\end{equation}
and the lemma follows by substituting (\ref{eq:42}) into (\ref{eq:40})
and simplifying.

\subsection{The Proof of Lemma~\ref{lemma:7}}
We start with
\begin{equation}
\label{eq:46}
\mding{182} = 
\frac{1}{L!} \int_{\R^L}
\det W^{\boldsymbol{\alpha}}_{\mathfrak{i}', \mathfrak{t}'} \cdot
\Pf T^{\boldsymbol{\alpha}}
\left\{\prod_{\ell=1}^L \phi(\alpha_{\ell})\right\} \,
d\lambda_L(\boldsymbol{\alpha}),
\end{equation}
where $\mf{t}$ is an element of $\mf{I}_{2M}^N$.
Expanding $\det W^{\boldsymbol{\alpha}}_{\mathfrak{i}',
\mathfrak{t}'}$ as a sum over $S_{L}$ allows us to write \ding{182} as
\begin{equation}
\label{eq:47}
\mding{182} = 
\frac{1}{L!} \sum_{\sigma \in S_L} \sgn(\sigma)
\underbrace{\int_{\R^L} \left\{ \prod_{\ell = 1}^L
    \phi(\alpha_{\ell}) \prod_{\ell=1}^L 
P_{\mathfrak{t}(\ell)}(\alpha_{\sigma(\ell)}) \right\} \;
\Pf T^{\boldsymbol{\alpha}} \, d\lambda_L(\boldsymbol{\alpha})}_{\mding{183}}.
\end{equation}$S_L$ naturally acts on $\R^L$ by permuting the
coordinates (denote this action by $\sigma \cdot \bs{\alpha}$), and it is easy to verify that for $\sigma \in S_L$, $\Pf T^{\sigma \cdot
\boldsymbol{\alpha}} = \sgn(\sigma) \Pf T^{\boldsymbol{\alpha}}$.
Using this fact, and the change of variables $\bs{\alpha} \mapsto \sigma^{-1} \cdot
\bs{\alpha}$ we may write \ding{183} as
\begin{eqnarray*}
\mding{183} = 
\sgn(\sigma^{-1}) \int_{\R^L} \left\{\prod_{\ell = 1}^L
\phi(\alpha_{\ell}) \prod_{\ell=1}^L
P_{\mathfrak{t}(\ell)}(\alpha_{\ell})\right\} \; \Pf T^{\boldsymbol{\alpha}}
\, d\lambda_L(\boldsymbol{\alpha}).
\end{eqnarray*}Substituting this into (\ref{eq:47}) we see that the sum over $S_L$
exactly cancels $1/L!$.  That is,
\begin{equation}
\label{eq:49}
\mding{182} = 
\int_{\R^L} \prod_{\ell=1}^L \left\{\phi(\alpha_{\ell}) \prod_{\ell =
  1}^L P_{\mathfrak{t}(\ell)}(\alpha_{\ell})\right\} \;
\Pf T^{\boldsymbol{\alpha}} \, d\lambda_{L}(\boldsymbol{\alpha}).
\end{equation}
Setting $K$
to the integer part of $(L+1)/2$, and using (\ref{eq:8}), we see
\begin{equation}
\label{eq:12}
\Pf T^{\bs{\alpha}} = \frac{1}{K!} \sum_{\tau \in \Pi_{2K}}
\sgn(\tau) \prod_{k=1}^K \sgn\left(\alpha_{\tau(2k)} - \alpha_{\tau(2k - 1)}\right).
\end{equation}
\subsubsection{$L$ Odd}

In the case of $L$ odd, we view $\alpha_{L+1} = +\infty$ so as to be
consistent with (\ref{eq:11}).  Substituting (\ref{eq:12}) into
(\ref{eq:49}) we find 
\begin{equation}
\label{eq:50}
\mding{182} = \frac{1}{K!} \sum_{\tau \in \Pi_{2K}} \sgn(\tau)
 \int_{\R^L} \underbrace{\prod_{\ell=1}^{L} \phi(\alpha_\ell) 
  P_{\mathfrak{t}(\ell)}(\alpha_{\ell}) \prod_{k=1}^K
  \sgn(\alpha_{\tau(2k)} - \alpha_{\tau(2k-1)} ) \,
  d\lambda_L(\boldsymbol{\alpha}) }_{\mding{184}}.
\end{equation}For each $\tau \in \Pi_{2K}$ there
is a $k_{\circ}$ such that $\alpha_{\tau(2 k_{\circ})} =
\alpha_{L+1}$.  If we set $\ell_{\circ} = {\tau(2 k_{\circ} -  1)}$
then we may write \ding{184} as
\begin{align*}
\mding{184} = \;& 
\phi(\alpha_{\ell_{\circ}})
P_{\mf{t}'(\ell_{\circ})}(\alpha_{\ell_{\circ}}) 
\Bigg\{\prod_{\ell = 1 \atop \ell \neq \ell_{\circ}}^L
\phi(\alpha_{\ell}) P_{\mf{t}'(\ell)}(\alpha_{\ell}) \prod_{k=1
  \atop k \neq k_{\circ}}^K \sgn(\alpha_{\tau(2k)} -
\alpha_{\tau(2k-1)})\Bigg\} \\
=\;& 
\phi(\alpha_{\ell_{\circ}})
P_{\mf{t}'(\ell_{\circ})}(\alpha_{\ell_{\circ}}) 
\Bigg\{
\prod_{k=1
  \atop k \neq k_{\circ}}^K \phi(\alpha_{\tau(2k)})
\phi(\alpha_{\tau(2k-1)}) \\
& \hspace{1cm} \times P_{(\mf{t}'\circ
  \tau)(2k)}(\alpha_{\tau(2k)}) P_{(\mf{t}'\circ
  \tau)(2k-1)}(\alpha_{\tau(2k-1)}) \sgn(\alpha_{\tau(2k)} -
\alpha_{\tau(2k-1)})\Bigg\},
\end{align*}
where the second equation comes from reindexing the first product by
$\ell \mapsto \tau^{-1}(\ell)$ together with the fact that $2(K-1) =
L-1$.  Substituting this into (\ref{eq:50}) and applying Fubini's
Theorem we find
\begin{align*}
\mding{182} =\;& \frac{1}{K!} \sum_{\tau \in \Pi_{2K}} \sgn(\tau)
\int_{\R} \phi(x) P_{(\mf{t}'\circ\tau)(2k_{\circ}-1)}(x) \, dx \\
& \hspace{1cm} \times \Bigg\{\prod_{k = 1 \atop k \neq k_{\circ}}^K \int_{\R^2}
\phi(x) \phi(y) P_{(\mf{t}'\circ\tau)(2k)}(y)
P_{(\mf{t}'\circ\tau)(2k-1)}(x) \sgn(y - x) \, dx \, dy\Bigg\} \\
=\;& \frac{1}{K!} \sum_{\tau \in \Pi_{2K}} \sgn(\tau) \left\{\int_{\R} \phi(x)
P_{(\mf{t}'\circ\tau)(2k_{\circ}-1)}(x) \, dx \right\}
\prod_{k=1 \atop k\neq k_{\circ}}^K
\la P_{(\mf{t}' \circ \tau)(2k-1)} , P_{(\mf{t}' \circ \tau)(2k)} \ra_{\R}
\end{align*}
Recalling the definition of $\mf{t}'_{\circ}$ gives $(\mf{t}'_{\circ} \circ \tau)(2k_{\circ}) = 2J$, and hence
\begin{equation}
\label{eq:51}
\mding{182} = \frac{1}{K!} \sum_{\tau \in \Pi_{2K}} \sgn(\tau) \;
  R_{\mf{t}'_{\circ}}[\tau(2 k_{\circ} - 1),\tau(2 k_{\circ})]
    \prod_{k=1 \atop k \neq k_{\circ}}^K R_{\mf{t}'_{\circ}}[\tau(2 k - 1),\tau(2 k)]
\end{equation}
\subsubsection{$L$ Even}
When $L$ is even, substituting (\ref{eq:12}) into (\ref{eq:49}) and simplifying, \ding{182} becomes
\begin{eqnarray}
\lefteqn{\frac{1}{K!} \sum_{\tau \in \Pi_{2K}} \sgn(\tau)
\Bigg\{\prod_{k = 1}^K \int_{\R^2}
\phi(x) \phi(y) P_{(\mf{t}'\circ\tau)(2k)}(y)
P_{(\mf{t}'\circ\tau)(2k-1)}(x) \sgn(y - x) \, dx \, dy\Bigg\}}
\nonumber \\
&& \hspace{1cm} =  \frac{1}{K!} \sum_{\tau \in \Pi_{2K}} \sgn(\tau) 
 \prod_{k=1}^K
\la P_{(\mf{t}' \circ \tau)(2k-1)} , P_{(\mf{t}' \circ \tau)(2k)} \ra_{\R}
. \hspace{3.5cm} \label{eq:53}
\end{eqnarray}

Regardless if $L$ is even or odd, (\ref{eq:53}) and
(\ref{eq:51}) imply that, 
\[
\mding{182} = \frac{1}{K!} \sum_{\tau \in \Pi_{2K}} \sgn(\tau)
\prod_{k=1}^K R_{\mf{t}'_{\circ}}[\tau(2 k - 1),\tau(2 k)] = \Pf R_{\mf{t}'_{\circ}}.
\]

\subsection{The Proof of Lemma~\ref{lemma:8}}
To prove Lemma~\ref{lemma:8} we set 
\[
\mding{172} = \frac{(-i)^M}{M!} \int\limits_{\C^M}
\left\{\prod_{m=1}^M \phi(\beta_m) \phi(\overline{\beta_m}) \sgn
\Im(\beta_m)\right\} \, \det W^{\bs{\beta}}_{\mf{i},\mf{t}} \;
d\lambda_{2M}(\bs{\beta}).
\]
From the definition of $W_{\mf{i}, \mf{t}}^{\bs{\beta}}$ we can write
\[
\det W_{\mf{i}, \mf{t}}^{\bs{\beta}} = \sum_{\tau \in S_{2M}}
\sgn(\tau) \prod_{m=1}^M P_{(\mf{t} \circ
  \tau)(2m-1)}(\overline{\beta_m}) P_{(\mf{t} \circ \tau)(2m)}(\beta_m).
\]
Substituting this into \ding{172} we see
\begin{eqnarray*}
\mding{172} &=& \frac{1}{M!} \sum_{\tau \in S_{2M}} \sgn(\tau) (-i)^M
\int_{\C^M} \left\{ 
\prod_{m=1}^M \phi(\overline{\beta_m}) \phi(\beta_m) \sgn
\Im(\beta_m) \right\} \\
&& \hspace{2cm} \times \left\{ 
\prod_{m=1}^M P_{(\mf{t} \circ
  \tau)(2m-1)}(\overline{\beta_m}) P_{(\mf{t} \circ \tau)(2m)}(\beta_m)
\right\} \, d\lambda_{2M}(\bs{\beta}).
\end{eqnarray*}
By Fubini's Theorem,
\begin{eqnarray*}
\mding{172} &=& \frac{1}{2^M M!} \sum_{\tau \in S_{2M}} \sgn(\tau) \Bigg\{
\prod_{m=1}^M (-2 i) \int_{\C} \phi(\overline{\beta})
\phi(\beta) \\
&& \hspace{4cm} \times P_{(\mf{t} \circ
  \tau)(2m-1)}(\overline{\beta}) P_{(\mf{t} \circ \tau)(2m)}(\beta)
\sgn \Im(\beta) d\lambda_2(\beta) \Bigg\} \\
&=& \frac{1}{2^M M!} \sum_{\tau \in S_{2M}} \sgn(\tau) \prod_{m=1}^M
\la P_{(\mf{t} \circ \tau)(2m-1)}, P_{(\mf{t} \circ \tau)(2m)} \ra_{\C},
\end{eqnarray*}
which is $\Pf C_{\mf{t}}$.  But, by definition, $\mf{t} =
\mf{t}_{\circ}$, and hence $\mding{172} = \Pf C_{\mf{t}_{\circ}}$ as
desired.
\subsection{The Proof of Corollary~\ref{cor:2}}

Corollary \ref{cor:2} follows from the fact that if $U$ is a $2J
\times 2J$ antisymmetric matrix, and $U[j,k] = 0$ when $j - k \equiv 0
\bmod 2$, then $\Pf U = \det A$ where $A$ is the $J \times J$ matrix
given by $A[j,k] = U[2j-1, 2k]$.  This is Lemma~8.10 of
\cite{sinclair-2005}, or can be proven by conjugating $U$ by an
appropriate permutation matrix $B$ so that
\[
B U B^{-1} = \begin{bmatrix}
0 & A \\
-A & 0
\end{bmatrix}.
\]
Then, using well-known results about the Pfaffian, $\Pf U = \Pf( B U
B^{-1}) = \det A$.

\section{Acknowledgements}

The author would like to thank Eugene Kanzieper and David Farmer for
many helpful suggestions.

\bibliography{bibliography}

\begin{center}
\vspace{.25cm}
\noindent\rule{4cm}{.5pt}

\vspace{.25cm}
{\small \noindent {\sc Pacific Institute for the Mathematical Sciences,} 
{\sc Vancouver, British Columbia} \\
email: {\tt sinclair@math.ubc.ca}}
\end{center}

\end{document}